\documentclass{article}
\usepackage[utf8]{inputenc}
\usepackage{epsfig}
\usepackage{graphicx}
\usepackage{amsbsy}
\usepackage{amsmath}
\usepackage{amsfonts}
\usepackage{amssymb}
\usepackage{textcomp}
\usepackage{hyperref}
\usepackage{aliascnt}
\usepackage{mathrsfs}

\newcommand{\mcm}[3]{\newcommand{#1}[#2]{{\ensuremath{#3}}}} 

\mcm{\tuple}{1}{\langle #1 \rangle}
\mcm{\name}{1}{\ulcorner #1 \urcorner}
\mcm{\Nbb}{0}{\mathbb{N}}
\mcm{\Zbb}{0}{\mathbb{Z}}
\mcm{\Rbb}{0}{\mathbb{R}}
\mcm{\Cbb}{0}{\mathbb{C}}
\mcm{\Qbb}{0}{\mathbb{Q}}
\mcm{\Acal}{0}{\cal A}
\mcm{\Bcal}{0}{\cal B}
\mcm{\Ccal}{0}{\cal C}
\mcm{\Dcal}{0}{\cal D}
\mcm{\Ecal}{0}{\cal E}
\mcm{\Fcal}{0}{\cal F}
\mcm{\Gcal}{0}{\cal G}
\mcm{\Hcal}{0}{\cal H}
\mcm{\Ical}{0}{\cal I}
\mcm{\Jcal}{0}{\cal J}
\mcm{\Kcal}{0}{\cal K}
\mcm{\Lcal}{0}{\cal L}
\mcm{\Mcal}{0}{\cal M}
\mcm{\Ncal}{0}{\cal N}
\mcm{\Ocal}{0}{{\cal O}}
\mcm{\Pcal}{0}{{\cal P}}
\mcm{\Qcal}{0}{{\cal Q}}
\mcm{\Rcal}{0}{{\cal R}}
\mcm{\Scal}{0}{{\cal S}}
\mcm{\Tcal}{0}{{\cal T}}
\mcm{\Ucal}{0}{{\cal U}}
\mcm{\Vcal}{0}{{\cal V}}
\mcm{\Wcal}{0}{{\cal W}}
\mcm{\Xcal}{0}{{\cal X}}
\mcm{\Ycal}{0}{{\cal Y}}
\mcm{\Zcal}{0}{{\cal Z}}
\mcm{\Mfrak}{0}{\mathfrak M}

\mcm{\restric}{0}{\upharpoonright}
\mcm{\upset}{0}{\uparrow}
\mcm{\onto}{0}{\twoheadrightarrow}
\mcm{\smallNbb}{0}{{\small \mathbb{N}}}
\DeclareMathOperator{\preop}{op}
\mcm{\op}{0}{^{\preop}}

\newcommand{\se}{\subseteq}
{\begin{array}{c}
\setlength{\unitlength}{1em}}%
{\end{array}}

\usepackage{amsthm}

\newcommand{\theoremize}[2]{\newaliascnt{#1}{thm} \newtheorem{#1}[#1]{#2} \aliascntresetthe{#1}}

\theoremstyle{plain}

\theoremize{lem}{Lemma}
\theoremize{skolem}{Skolem}
\theoremize{fact}{Fact}

\theoremize{obs}{Observation}
\theoremize{prop}{Proposition}
\theoremize{cor}{Corollary}
\theoremize{que}{Question}
\theoremize{oque}{Open Question}
\theoremize{oprob}{Open Problem}
\theoremize{con}{Conjecture}
\theoremize{setting}{Setting}

\theoremstyle{definition}
\theoremize{dfn}{Definition}

\theoremize{rem}{Remark}
\theoremize{eg}{Example}
\theoremize{exercise}{Exercise}
\theoremstyle{plain}

\usepackage{graphicx, hyperref, xcolor, nicefrac}
\usepackage{subcaption}
\usepackage{csquotes,amsmath}
\usepackage[british]{babel}
\usepackage{tabularx}
\usepackage{algorithm}
\usepackage{algorithmic}
\usepackage{lscape}
\usepackage{verbatim}

\newcommand{\algorithmicfunction}{\textbf{function}}
\newcommand{\algorithmicendfunction}{\algorithmicend\ \algorithmicfunction}
\makeatletter
\newcommand\FUNCTION[3][default]{%
\ALC@it \algorithmicfunction\ \textsc{#2}(#3)%
 \ALC@com{#1}%
\begin{ALC@prc}%
}
\newcommand\ENDFUNCTION{%
  \end{ALC@prc}%
  \ifthenelse{\boolean{ALC@noend}}{}{%
    \ALC@it\algorithmicendfunction
  }%
}
\newenvironment{ALC@prc}{\begin{ALC@g}}{\end{ALC@g}}
\makeatother

\title{How Local Separators Shape Community Structure in Large Networks}
\author{Sarah Frenkel and Johannes Carmesin}
 \date{\today}

\begin{document}

\maketitle

\begin{abstract}
Community detection is a key tool for analyzing the structure of large networks. Standard methods, such as modularity optimization, focus on identifying densely connected groups but often overlook natural local separations in the graph. In this paper, we investigate \emph{local separator methods}, which decompose networks based on structural bottlenecks rather than global connectivity. We systematically compare them with well-established community detection algorithms on large real-world networks. Our results show that \emph{local 1-separators} consistently identify the densest communities, outperforming modularity-based methods in this regard, while \emph{local 2-separators} reveal hierarchical structures but may over-fragment small clusters. These findings are particularly strong for road networks, suggesting practical applications in transportation and infrastructure analysis. Our study highlights local separators as a scalable and interpretable alternative for network decomposition.
\end{abstract}

\section{Introduction}

Community detection is a fundamental task in network science, commonly approached using modularity-based methods~\cite{newman2004finding, fortunato2010community}. While these methods have been widely used, they often struggle to capture fine-grained structures in networks where \emph{local bottlenecks} define natural divisions rather than global connectivity patterns.

\begin{figure}
    \centering
        \includegraphics[width=0.5\textwidth]{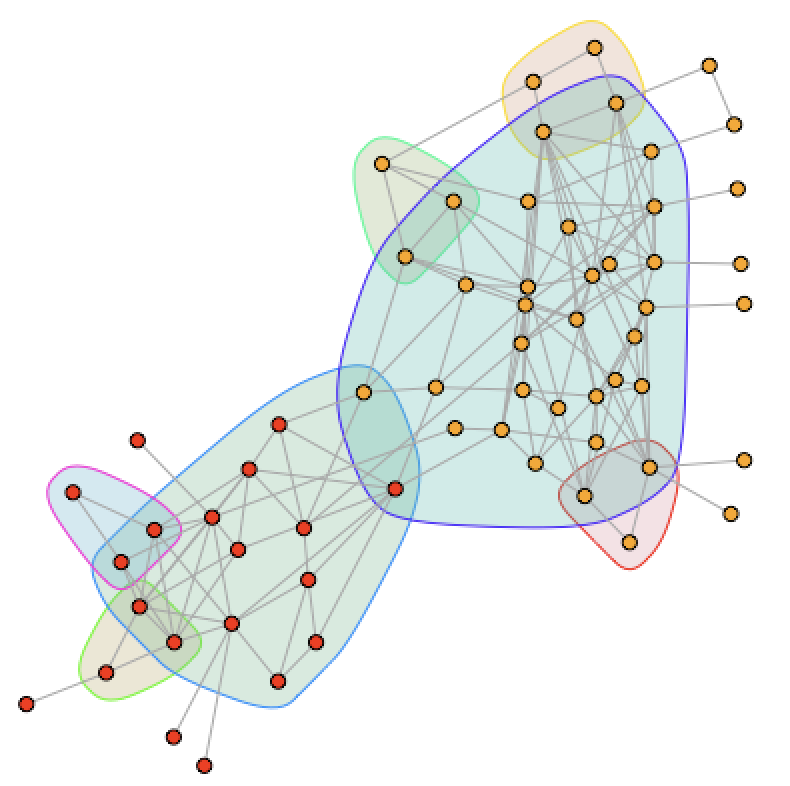}
        \caption{The ground truth for this network is that the dolphins are partitioned into two  social groups, the red nodes and the orange ones. The clusters of our decomposition method are highlighted by the coloured regions. They are able to reconstruct this partition up to two nodes at the boundary that are assigned to both clusters (as we allow an overlapping clustering), and it also provides a finer structure within the red and orange families.}\label{fig:dol}

    \end{figure}

A simple example illustrating this issue is the well-studied \emph{dolphins social network}~\cite{lusseau2003bottlenose}, where the ground truth communities are known, see \autoref{fig:dol}. Standard modularity-based algorithms only partially recover the expected structure, whereas \emph{local separator methods}, which decompose a network based on its natural local separations, successfully identify most of the known communities. This motivates the question of how local separators perform in larger real-world networks.

Local separators were recently introduced as a way to decompose large networks based on local bottlenecks rather than global connectivity. Roughly speaking, a local separator is a small set of vertices that separates its local neighbourhood into well-connected regions. This allows for a natural and scalable decomposition into so-called bags—components separated by these bottlenecks—which often correspond to meaningful communities. The concept was originally developed in~\cite{carmesin2022local} as a theoretical tool to generalise ideas from tree decompositions. A simpler and practically usable definition, along with the first algorithm to compute such decompositions, was introduced in \cite{carmesin2023how}.
Since local separators rely only on neighbourhood-level computations, the method scales naturally to large sparse graphs—making it practical even with minimal implementation overhead.

In this paper, we test local separator methods on large networks without an assumed ground truth, including collaboration and transportation networks. Our empirical results show that:
\begin{itemize}
    \item The \emph{local 1-separator method} consistently finds the densest clusters, outperforming modularity-based methods in this regard.
    \item The \emph{local 2-separator method} provides a finer decomposition, revealing hierarchical structure but sometimes over-fragmenting small clusters.
    \item The results are particularly strong on \emph{road networks}, suggesting that local separators are well-suited for transportation and infrastructure analysis.
\end{itemize}

These findings suggest that local separator methods offer a practical and scalable alternative to traditional community detection algorithms, particularly in settings where structural separations are more relevant than modularity optimization.

The remainder of the paper is structured as follows. In \autoref{sec:comparison}, we outline the methodology used to compare community detection algorithms. In \autoref{sec:netscience}–\autoref{sec:nrw}, we present empirical results on different network datasets. We conclude with a discussion of our findings in \autoref{sec:conclusion}.

\section{How to Compare Community Detection Algorithms?}\label{sec:comparison}

According to Wikipedia, \lq The evaluation of algorithms, to detect which are better at detecting community structure, \lq is still an open question\rq~\cite{enwiki:1223538653}. While this remains an open problem, in this section, we systematically compare our new algorithm with widely used community detection methods from the literature. Recognizing that no single evaluation metric is universally accepted, we adopt multiple perspectives to assess performance. 

The remainder of this section is structured as follows. In the first subsection, we introduce \lq modularity\rq\ and \lq overlapping modularity\rq, two widely used measures for evaluating community detection algorithms. In the second subsection, we describe the community detection algorithms used in our comparisons. 

In the third subsection, we evaluate the performance of our algorithm against existing methods using overlapping modularity as a benchmark. However, in the fourth subsection, we argue that overlapping modularity alone is insufficient, particularly in our case, and explain why additional evaluation criteria are necessary.  

Finally, in the fifth subsection, we propose a new approach for comparing community detection algorithms, specifically tailored to our context. In the following sections, we apply this measure to compare our community detection algorithms via local separators with standard methods from the literature.

\subsection{An Introduction to Modularity}

A widely used measure for comparing community detection algorithms is \lq{modularity}\rq~\cite{fortunato2010community,newman2004finding}, or \lq{overlapping modularity}\rq\ when communities are allowed to overlap~\cite{chen2014extension}. Overlapping modularity can be viewed as a {fractional relaxation} of standard modularity and is defined as follows.

Given two vertex sets $A$ and $B$, we denote by $P(A,B)$ the pairs $(i,j)$ with $i\in A$ and $j\in B$.
Given a graph \( G = (V, E) \) and a set  \( C\se 2^{V(G)} \), which consists of (potentially) overlapping vertex sets and whose elements $c\in C$ are referred to as \emph{communities}, the \emph{overlapping modularity} is given by:

\[
Q_{ov} = \sum_{c\in C} \left[ \frac{|E^{in}_c|}{|E|} - \left(\frac{2|E^{in}_c| + |E^{out}_c|}{ 2|E|} \right)^2\right]
\]

where $|E^{in}_c| = \frac{1}{2}\sum\limits_{(i,j) \in P(c,c)}a_{i,c}a_{j,c}$ and $|E^{out}_c| = \sum_{i\in C} \sum\limits_{d \in C \atop d\neq c}\sum_{(i,j)\in P(c,d)}a_{i,c}a_{j,d}$. The values $a_{i,c}$ measure how strong a vertex $i$ belongs to a community $c$ and is defined as  $a_{i,c} = \frac{1}{|\{d\in C| i\in d\}|}$ if $i\in c$, and zero otherwise.

It follows that computing the overlapping modularity for a \emph{non-overlapping} community structure yields the same result as the standard (non-overlapping) modularity.

One major advantage of (overlapping) modularity is that it maps a set of communities \( C \) in a graph to a single real number in \( (-\infty,1] \). This provides a simple and intuitive way to compare different community detection algorithms: the higher the modularity value, the better the algorithm performs on this particular instance.

\subsection{Community Detection Methods}

A variety of approaches exist for detecting community structure in networks, each with different underlying principles. 
For a more detailed account, we refer to the surveys~\cite{fortunato2010community, javed2018community}.

The problem is to decompose a graph into communities (or clusters), which are vertex sets with only a few ties to the rest of the network. Different methods interpret \enquote{few ties} in different ways. 

\emph{Modularity-based algorithms}~\cite{newman2004finding} optimize a global objective function that favors densely connected groups but are known to suffer from resolution limits~\cite{fortunato2007resolution}, often failing to detect smaller communities.  

Several methods leverage \emph{random walks} to infer community structure. \emph{Infomap}~\cite{rosvall2008maps} detects communities by modeling information flow and partitioning nodes based on how likely a walker remains within the same region of the graph. A related approach, \emph{Flow Stability}~\cite{vanderhoorn2015flow}, generalizes this idea by analyzing how long a random walker is retained in a cluster over multiple time scales, allowing for multi-resolution detection of communities.

\emph{Spectral clustering methods}~\cite{von2007tutorial} use eigenvector decompositions to infer partitions, but their results can be sensitive to small perturbations in network structure.  
\emph{Label propagation algorithms}~\cite{raghavan2007near} offer a computationally efficient but heuristic-based alternative, assigning nodes iteratively based on neighboring labels.  
\emph{Betweenness-based methods}, such as the \emph{Girvan-Newman algorithm}~\cite{newman2004finding}, remove edges iteratively based on their betweenness centrality to separate clusters, but this approach does not scale well to large graphs.  

Beyond these classical methods, tree-based and structural decomposition approaches have been explored. 
\emph{Tree decompositions} were used by Adcock, Mulligan, and Sullivan~\cite{adcock2013tree} to detect community structure, though their approach lacked scalability.  
\emph{Tangles}, a concept from graph minor theory, have been employed for clustering by Klepper et al.~\cite{klepper2023clustering}, offering another structural perspective.  

We compare our algorithm with four widely used community detection algorithms that are readily available in the igraph Python library~\cite{csardi2006igraph}. These methods were selected as they represent a diverse set of approaches:

\begin{itemize}
    \item \textbf{Infomap (IM)}: Based on \emph{information flow} and random walks, this method captures \emph{network dynamics} rather than just structural properties~\cite{rosvall2008maps, rosvall2009map}.
    \item \textbf{Label Propagation (LP)}: A fast and \emph{heuristic} method that identifies communities without optimizing a specific objective function~\cite{raghavan2007near}.
    \item \textbf{Leiden Modularity (LM)}: A \emph{modularity-based} approach that improves upon the widely used Louvain method, ensuring more stable partitions~\cite{traag2019louvain}.
    \item \textbf{Best Multi-Level (BML)}: Another \emph{hierarchical modularity-based} approach that efficiently detects community structure in large networks~\cite{blondel2008fast}.
\end{itemize}

This selection ensures that we compare our method against \emph{both modularity-based and alternative approaches}, providing a broad perspective on how different strategies perform.

\subsection{Comparing Performances Based on Overlapping Modularity}

In this subsection, we compare the overlapping modularity of the communities detected by different community detection algorithms, including our $1$- and $2$-separator methods. We apply these algorithms to the networks listed in \autoref{tab:networks}.

\begin{table}[htbp]
\centering
\begin{tabular}{|l|c|c|l|}
\hline
Name & $n$ & $m$ & Description \\
\hline
Dolphins & $62$ & $159$ & Social network of bottlenose dolphins~\cite{konect:2017:dolphins} \\
Euroroads & $1174$ & $1417$ & Road network connecting cities in Europe~\cite{konect:2017:subelj_euroroad} \\
Netscience & $1461$ & $2742$ & Co-authorship network in the field of network science~\cite{konect:2018:dimacs10-netscience} \\
Powergrid & $4941$ & $13188$ & Power grid of parts of the United States~\cite{konect:2017:opsahl-powergrid} \\
Nrw & $9133$ & $14125$ & Road network of parts of Germany \\
\hline
\end{tabular}
\caption{Description of the networks used in our experiments.}
\label{tab:networks}
\end{table}

The results of our modularity-based evaluation are summarized in \autoref{modularity23}, where we report both the number of detected communities ($|B|$) and the modularity scores obtained by each method.

\begin{table}[htbp]
\centering
\begin{tabular}{|c|c|c|c|c|c|c|c|c|c|c|}
\hline
\multicolumn{3}{|c|}{Network} & \multicolumn{3}{c|}{Local 1-separator} & \multicolumn{3}{c|}{Local 2-separator} & \multicolumn{2}{c|}{Infomap} \\ 
Name & $n$ & $m$ & $d$ & $|B|$ & Ov Mod & $d$ & $|B|$ & Ov Mod & $|B|$ & Mod \\
\hline
Dolphins & 62 & 159 & 3 & 4 & 0.4632 & 4 & 16 & 0.3592 & 5 & 0.5277 \\
Euroroads & 1174 & 1417 & 6 & 27 & 0.5355 & 9 & 85 & 0.3697 & 160 & 0.7880 \\
Netscience & 1461 & 2742 & 5 & 124 & 0.8779 & 7 & 155 & 0.8122 & 314 & 0.9299 \\
Powergrid & 4941 & 13188 & 7 & 95 & 0.7274 & 14 & 145 & 0.7232 & 1785 & 0.4762 \\
Nrw & 9133 & 14125 & 10 & 567 & 0.9002 & 17 & 2246 & 0.6273 & 911 & 0.8631 \\
\hline
\end{tabular}

\vspace{0.3cm}

\begin{tabular}{|c|c|c|c|c|c|c|c|c|}
\hline
\multicolumn{3}{|c|}{Network} & \multicolumn{2}{c|}{Label Propagation} & \multicolumn{2}{c|}{Best Multilevel} & \multicolumn{2}{c|}{Leiden (mod)} \\ 
Name & $n$ & $m$ & $|B|$ & Mod & $|B|$ & Mod & $|B|$ & Mod \\
\hline
Dolphins & 62 & 159 & 5 & 0.5047 & 5 & 0.5233 & 5 & 0.5241 \\
Euroroads & 1174 & 1417 & 118 & 0.8124 & 46 & 0.8802 & 47 & 0.8863 \\
Netscience & 1461 & 2742 & 330 & 0.9104 & 277 & 0.9587 & 279 & 0.9594 \\
Powergrid & 4941 & 13188 & 503 & 0.7999 & 42 & 0.9353 & 41 & 0.9388 \\
Nrw & 9133 & 14125 & 1458 & 0.7799 & 62 & 0.9523 & 66 & 0.9558 \\
\hline
\end{tabular}
\caption{Modularity scores of decompositions based on local 1- and 2-separators. 
The number of detected communities, denoted by $|B|$, is also shown for each method. 
As expected, methods explicitly optimizing modularity, such as \emph{Best Multilevel} and \emph{Leiden (mod)}, achieve the highest modularity scores. 
Local separators, while not designed to maximize modularity, still yield comparable scores to \emph{Infomap} and \emph{Label Propagation}.}
\label{modularity23}
\end{table}

\subsubsection*{Interpreting Modularity Scores}

Modularity is a widely used metric in community detection, designed to measure how well a given partition captures densely connected groups while minimizing inter-community connections~\cite{newman2004finding}. Since it is explicitly optimized by certain algorithms, it is expected that methods such as \emph{Best Multilevel} and \emph{Leiden (mod)}, which maximize modularity, achieve the highest scores in this measure. 

While our approach based on local separators is not designed to optimize modularity, we observe that it still achieves scores that are comparable to other well-established algorithms such as \emph{Infomap} and \emph{Label Propagation}, which do not explicitly maximize modularity either. This suggests that local separators identify meaningful structural divisions in the network, even when evaluated through a modularity-based lens.

\subsection{Limitations of Modularity}

Modularity is a widely used benchmark in network science; however, it has well-known limitations that make it less suitable in certain contexts. A key issue is its tendency to favor partitions with a smaller number of large communities~\cite{fortunato2007resolution}, often overlooking finer structural separations. This is particularly problematic in networks where \emph{local bottlenecks} define natural divisions that modularity-based methods fail to capture.  

For example, if all nodes are placed in a single community ($C = \{G\}$), modularity reaches its maximum possible value of one. Similarly, given a set of communities $C = \{c_1, \dots, c_n\}$, the modularity of the modified set $\{c_1 \cup c_2, c_3, \dots, c_n\}$ is not smaller than that of $C$. This suggests that when comparing algorithms that produce similar numbers of communities, modularity may provide a reasonable measure of quality. However, when comparing algorithms that generate vastly different numbers of communities, the observed modularity differences may be largely influenced by the number of detected communities rather than the \emph{intrinsic quality} of the community structure itself.

A similar issue arises when comparing two sets of detected communities, $C_1$ and $C_2$, that have different distributions of community sizes. If one community contains nearly all the vertices in the network, modularity is artificially inflated, often approaching one. Consequently, differences in modularity values across algorithms may stem more from the number and size distribution of the detected communities rather than from meaningful differences in how well they capture the network’s underlying \emph{community structure}.

While modularity has been widely used and produces meaningful results in some cases~\cite{newman2006modularity}, these issues suggest that relying on it as the \emph{sole criterion} for evaluating community detection is problematic. A more robust evaluation should consider additional structural properties beyond modularity alone.

\subsection{Community Detection Problem}

To compare the performance of different community detection algorithms, a natural approach is to first define the desired properties of a community and then simply count how many such communities each algorithm identifies. The more communities an algorithm finds, the better its performance. However, if we define a notion of community that is tailored specifically to our algorithm, then it will, by definition, perform exceptionally well. To avoid this bias, we adopt a more general and algorithm-independent definition of a community.

Our approach is to define a \emph{density function} for each candidate community $c$ in a set $C$ of vertex subsets of a graph. A subset $c$ is considered a community if its density exceeds a given threshold. In graph theory, there are two widely used ways to measure density, depending on the graph type:

\begin{itemize}
    \item In \emph{dense graphs}, where the number of edges is quadratic in the number of vertices, density is typically defined as $\frac{e(c)}{|c|^2}$, where $e(c)$ denotes the number of edges within $c$.
    \item In \emph{sparse graphs}, where the number of edges is proportional to the number of vertices, as is the case for the planar networks studied in this paper, density is commonly defined as $\frac{e(c)}{|c|}$.
\end{itemize}

To define density in the context of a graph $G$ with a set $C$ of potential communities, we utilize the belonging coefficients $a_{i,c}$ introduced earlier. Instead of using $e(c)$ directly, we replace it with 
\[
\frac{1}{2} \sum\limits_{(i,j) \in P(c,c)} a_{i,c} a_{j,c},
\]
where $P(c,c)$ denotes the pairs of endvertices of edges with both endpoints in $c$. This yields the following formula for the density $\beta(c)$ of $c$:

\[
\beta(c) = \frac{1}{2 |c|} \sum\limits_{(i,j) \in P(c,c)} a_{i,c} a_{j,c}.
\]

The following example demonstrates why it is not meaningful to assume the existence of a universal constant $\delta$ such that, for every graph $G$ and every set $C$, a subset $c \in C$ is a community if and only if $\beta(c) \geq \delta$.

\begin{eg}\label{no-d}
Let $\delta$ be an absolute constant. Suppose we are given a dataset with a ground truth specifying the actual communities. Further, assume that this dataset can be represented as a graph $G$ in such a way that the true communities correspond exactly to a set $C$ of vertex subsets, all of which have density above $\delta$.

Now, construct a new graph $H$ from $G$ by adding a duplicate of each vertex $v$ and connecting it to the neighbors of $v$ in $G$ as well as the duplicates of those neighbors. The communities in $H$ should be obtained from those in $G$ by including, for each vertex $v \in c$, its duplicate. However, each such community in $H$ has twice the density of the corresponding community in $G$. Thus, for $H$, a reasonable threshold would be $2\delta$ instead of $\delta$.
\end{eg}

Given \autoref{no-d}, it is natural to define a \emph{community detection problem} as follows: A community detection problem consists of a graph $G$ together with a parameter $\delta$, and the goal is to find a set $C$ of vertex subsets of $G$ that is as large as possible while ensuring that $\beta(c) \geq \delta$ for all $c \in C$.

\section{Empirical Analysis on the Netscience Network}\label{sec:netscience}

The Netscience network is a collaboration network among scientists with $n=1461$ vertices and $m=2742$ edges~\cite{newman2006finding}. 

\begin{figure}
\makebox[\textwidth][c]{
\begin{subfigure}[t]{0.8\textwidth}
     \centering        \includegraphics[width=\columnwidth]{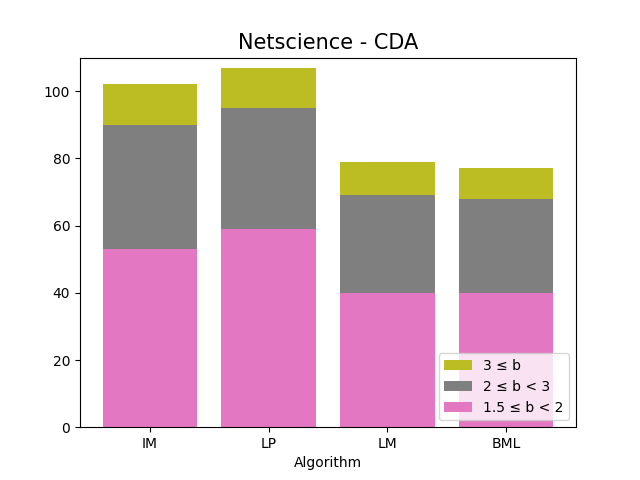} 
    \caption{Community detection algorithms (IM, LP, LM, BML)}
\end{subfigure}
    \hfill
\begin{subfigure}[t]{0.8\textwidth}
     \centering        \includegraphics[width=\columnwidth]{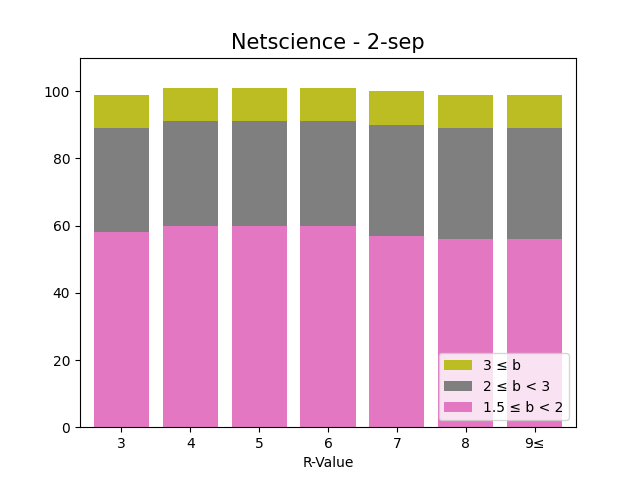} 
    \caption{Local 2-separator method}
    \end{subfigure}}

\centering

\begin{subfigure}[t]{0.8\textwidth}
     \centering        \includegraphics[width=\columnwidth]{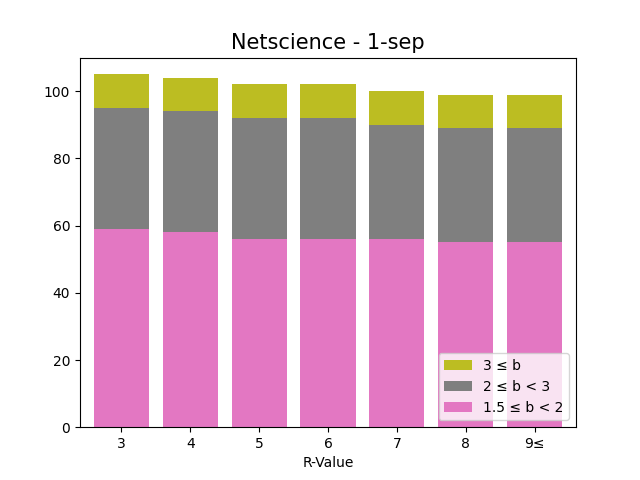} 
    \caption{Local 1-separator method}
\end{subfigure}
        \caption{Comparison of community detection methods applied to the Netscience network. 
        (a) Results from Infomap (IM), Label Propagation (LP), Leiden Modularity (LM), and Best Multilevel (BML). 
        (b) Results from the local 1-separator method. 
        (c) Results from the local 2-separator method.}\label{netscience}
\end{figure}

\autoref{netscience} compares standard community detection algorithms (IM, LP, LM, BML) with our local separator methods on this dataset. IM and LP yield nearly identical numbers of clusters, indicating that they provide stable results on this network. In contrast, modularity-based methods (LM, BML) identify fewer but not necessarily denser clusters, suggesting they emphasize larger, more globally distinct communities while overlooking finer structures.

The local 1-separator and 2-separator methods find a comparable number of clusters to IM and LP, reinforcing the idea that this network contains well-defined community structure. The agreement across methods suggests that the network’s clustering patterns are robust, regardless of the computational approach used.

\section{Empirical Analysis on the Euroroads Network}\label{sec:euroroads}

The Euroroads network consists of $n = 1174$ vertices, where each vertex represents a major city in Europe, and $m = 1417$ edges, representing unweighted connections between them~\cite{vsubelj2011robust}. Unlike real-world road networks, the absence of edge weights means that distances and travel times are not encoded.

\begin{figure}
\makebox[\textwidth][c]{
\begin{subfigure}[t]{0.8\textwidth}
     \centering        \includegraphics[width=\columnwidth]{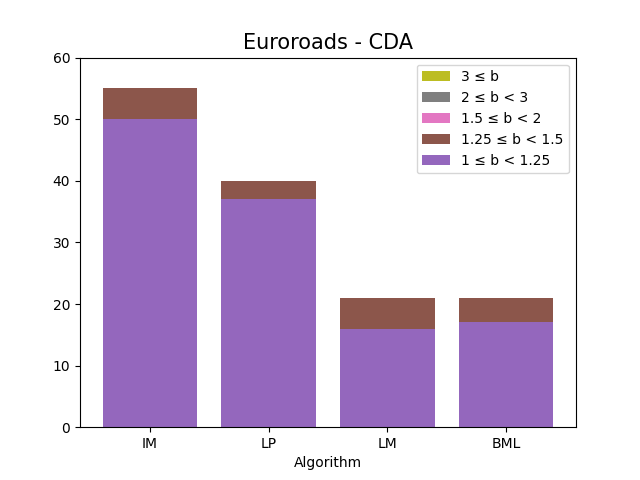} 
    \caption{Community detection algorithms (IM, LP, LM, BML)}
\end{subfigure}
    \hfill
\begin{subfigure}[t]{0.8\textwidth}
     \centering        \includegraphics[width=\columnwidth]{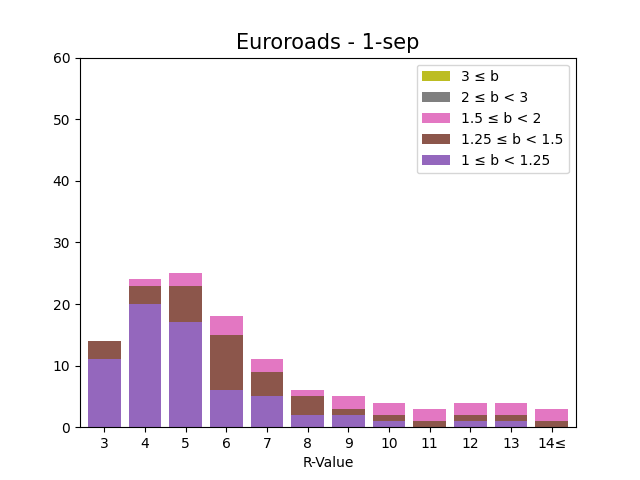} 
    \caption{Local 1-separator method}
\end{subfigure}}
\centering
\begin{subfigure}[t]{0.8\textwidth}
     \centering        \includegraphics[width=\columnwidth]{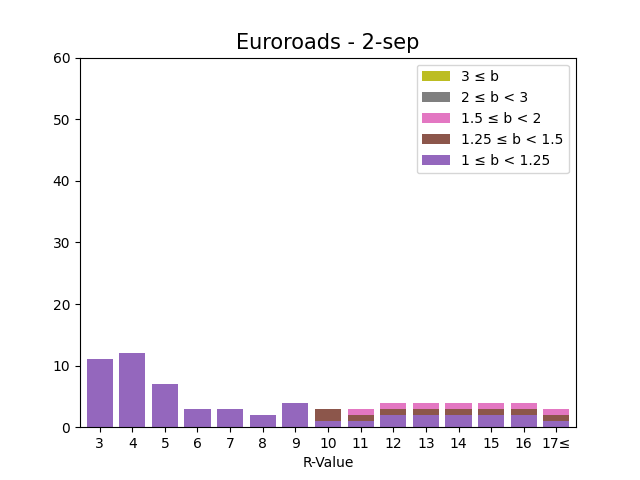} 
    \caption{Local 2-separator method}
\end{subfigure}
        \caption{Comparison of community detection methods applied to the Euroroads network. 
        (a) Infomap (IM), Label Propagation (LP), Leiden Modularity (LM), and Best Multilevel (BML). 
        (b) Local 1-separator method. 
        (c) Local 2-separator method.}
\label{euroroads}
\end{figure}

\autoref{euroroads} compares standard community detection algorithms (IM, LP, LM, BML) with our local separator methods. IM and LP detect more clusters, while modularity-based methods (LM, BML) find fewer but not necessarily denser ones. 

The \emph{local 1-separator method} consistently produces the \emph{densest clusters}, outperforming all other methods in this respect. In contrast, the \emph{local 2-separator method} decomposes the network too finely, leading to over-fragmentation, particularly in smaller clusters. 

These results suggest that local separators, particularly the \emph{1-separator method}, are well-suited for detecting structural regions in road networks. The similarities between this dataset and the NRW network (next section) indicate that our methods generalize well to transportation networks.

\section{Empirical Analysis on the NRW Road Network}\label{sec:nrw}

The road network of the German state of North Rhine-Westphalia (NRW) consists of $n = 9133$ vertices and $m = 14125$ edges. Unlike the Euroroads network, where each edge represents a direct connection between major cities, the NRW network includes additional \emph{subdivision points along long roads}, providing a more detailed representation of the infrastructure.

To improve the meaningfulness of detected communities, we applied preprocessing: vertices of degree 1 were removed, and vertices of degree 2 were replaced by merging their incident edges into a single edge.

\begin{figure}
\makebox[\textwidth][c]{
\begin{subfigure}[t]{0.8\textwidth}
     \centering        \includegraphics[width=\columnwidth]{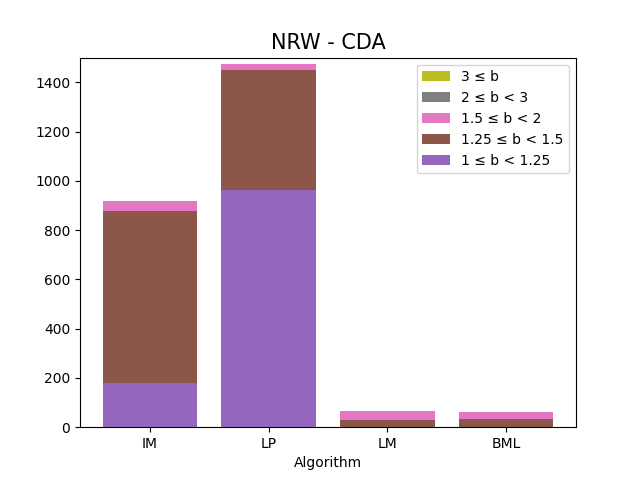} 
    \caption{Community detection algorithms (IM, LP, LM, BML)}
\end{subfigure}
    \hfill
\begin{subfigure}[t]{0.8\textwidth}
     \centering        \includegraphics[width=\columnwidth]{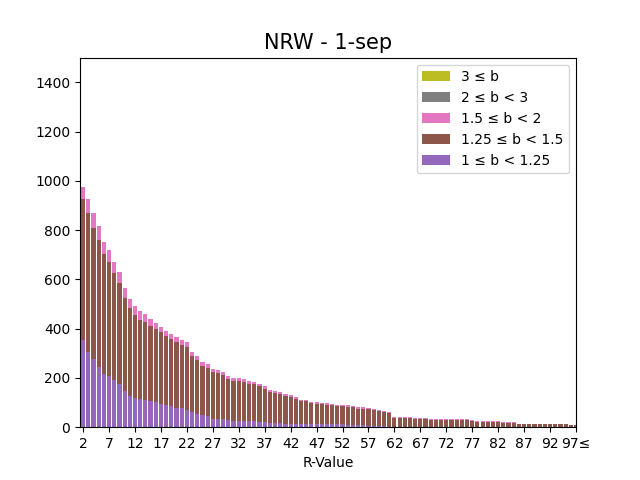} 
    \caption{Local 1-separator method}
\end{subfigure}}
\centering
\begin{subfigure}[t]{0.8\textwidth}
     \centering        \includegraphics[width=\columnwidth]{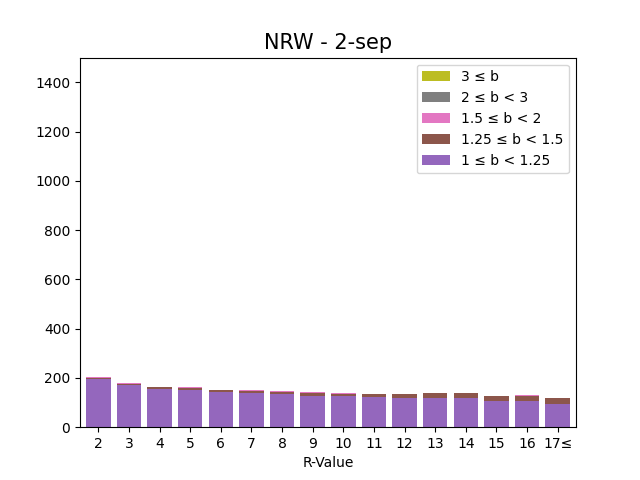} 
    \caption{Local 2-separator method}
\end{subfigure}
        \caption{Comparison of community detection methods applied to the NRW road network. 
        (a) Infomap (IM), Label Propagation (LP), Leiden Modularity (LM), and Best Multilevel (BML). 
        (b) Local 1-separator method. 
        (c) Local 2-separator method.}
\label{nrw}
\end{figure}

\autoref{nrw} compares standard community detection algorithms (IM, LP, LM, BML) with our local separator methods. The overall findings align closely with those from the Euroroads network, reinforcing that our methods generalize well to transportation networks.

As in the Euroroads network, IM and LP detect a higher number of clusters, while modularity-based methods (LM, BML) produce fewer but not necessarily denser communities. The \emph{local 1-separator method} consistently identifies the \emph{densest clusters}, outperforming all other methods. 

Interestingly, \emph{local 2-separators} reveal \emph{hierarchical substructures} within larger communities~\cite{cf_locsepr-algo}, making them more effective in detecting multi-scale structures. However, they may \emph{over-fragment} small clusters, potentially dissolving meaningful units. 
A similar phenomenon was observed in the \emph{dolphins social network} discussed in the Introduction.

The consistency of these results across multiple road networks suggests that local separator methods provide a robust and scalable approach for transportation network analysis. The \emph{local 1-separator method} identifies the densest clusters, efficiently capturing well-defined communities. However, the \emph{local 2-separator method} plays a crucial role in uncovering finer substructures within these dense regions. While splitting large, high-density communities is inherently more challenging, local 2-separators enable a \emph{second layer of refinement}, revealing hierarchical organization that other methods fail to detect. This layered approach—first applying local 1-separators to identify major divisions, followed by local 2-separators to refine large clusters—demonstrates the flexibility and depth of local separator methods in network analysis~\cite{cf_locsepr-algo}.

\section{Concluding Remarks}\label{sec:conclusion}

In this work, we have demonstrated that \emph{local separator methods} provide a practical and effective approach for detecting meaningful structure in large networks. Our empirical analysis across multiple datasets, particularly road networks, highlights their ability to decompose graphs into dense, well-structured clusters.  

Among the methods tested, the \emph{local 1-separator method} consistently identified the densest communities while maintaining a reasonable number of clusters, outperforming standard modularity-based methods in this regard. The \emph{local 2-separator method}, while sometimes over-fragmenting smaller clusters, revealed hierarchical substructures when applied to sufficiently large communities, making it a useful tool for multi-scale network analysis.

Future research could explore the applicability of local separators beyond road networks, assessing their effectiveness in other sparse graphs, such as biological, social, and communication networks. Integrating local separator methods with existing community detection frameworks may also provide a complementary approach, balancing modularity optimization with structural decomposition.  

Local separators have potential applications in \emph{urban mobility analysis}, \emph{biological network modelling}, and \emph{infrastructure resilience studies}~\cite{rosvall2008maps, evans2009line}. Further exploration in these areas could lead to refinements and extensions tailored to domain-specific challenges, advancing both theoretical and practical understanding of network decomposition techniques.

\bibliographystyle{plain}
\bibliography{literatur.bib}
\end{document}